\documentclass[conference]{IEEEtran}
\usepackage{graphicx}
\usepackage{algorithm}
\usepackage[noend]{algorithmic}
\usepackage{subfig}
\usepackage{amssymb,amsmath,graphicx,charter,latexsym,cite,url}
\usepackage{color}

\newtheorem{theorem}{\textbf{Theorem}}

\renewcommand{\IEEEQED}{\IEEEQEDopen}

\begin{document}
\title{An Energy-Aware Protocol for Self-Organizing Heterogeneous LTE Systems}
\author{
\IEEEauthorblockN{I-Hong Hou$^{*}$ and Chung~Shue~Chen$^{\dag}$}
\IEEEauthorblockA{
$^{*}$Computer Engineering and Systems Group \& ECE, Texas A\&M University, TX 77843, USA\\
$^{\dag}$Alcatel-Lucent Bell Labs, Centre de Villarceaux, 91620 Nozay, France\\
Email: ihou@tamu.edu, cs.chen@alcatel-lucent.com}
}

\maketitle

\begin{abstract}
This paper studies the problem of self-organizing heterogeneous LTE systems. We propose a model that jointly considers several important characteristics of heterogeneous LTE system, including the usage of orthogonal frequency division multiple access (OFDMA), the frequency-selective fading for each link, the interference among different links, and the different transmission capabilities of different types of base stations. We also consider the cost of energy by taking into account the power consumption, including that for wireless transmission and that for operation, of base stations and the price of energy. 
Based on this model, we aim to propose a distributed protocol that improves the spectrum efficiency of the system, which is measured in terms of the weighted proportional fairness among the throughputs of clients, and reduces the cost of energy. We identify that there are several important components involved in this problem. We propose distributed strategies for each of these components. Each of the proposed strategies requires small computational and communicational overheads. Moreover, the interactions between components are also considered in the proposed strategies. Hence, these strategies result in a solution that jointly considers all factors of heterogeneous LTE systems. Simulation results also show that our proposed strategies achieve much better performance than existing ones.
\end{abstract}

\begin{IEEEkeywords}
Self-organizing networks, LTE, OFDMA, proportional fairness, energy effieincy.
\end{IEEEkeywords}

\section{Introduction}  
\label{section:introduction}

With the foreseen exponentially increasing number of users and traffic in 4G and LTE/LTE-Advanced (LTE-A) systems  \cite{3GPP-LTE-A}, existing deployment and practice of cellular radio networks that strongly rely on highly hierarchical architectures with centralized control and resource management becomes economically unsustainable. Network self-organization and self-optimization are among the key targets of future cellular networks so as to relax the heavy demand of human efforts in the network planning and optimization tasks and to reduce the system's capital and operational expenditure (CAPEX/OPEX) \cite{BLTJ10}.

The next-generation mobile networks are expected to provide a full coverage of broadband wireless service and support fair and efficient resource utilization with a high degree of operation autonomy and system intelligence \cite{NGMN}. In addition, energy efficiency has emerged as an important concern for future mobile networks. It is expected that energy consumption by the information and communications technology (ICT) industry will be rising at $15-20\%$ per year \cite{GF08}, and hence energy bills will become an important portion of operational expenditure. To reduce the impacts on both revenue and environment caused by energy consumption, while providing satisfactory services to customers, a mechanism that jointly improves spectrum and energy efficiency for self-organizing networks is needed.

In this paper, we study the problem of self-organizing heterogeneous LTE systems and aim to achieve both spectrum and energy efficiency. We propose a generic model that jointly takes into account the key characteristics of today's LTE networks \cite{LTEbook2011}, including the usage of orthogonal frequency division multiple access (OFDMA) in the air interface, the nature of frequency-selective fading for each link, 
multi-cell multi-link interference occured, 
and the different transmission (power) capabilities of different types of base stations,
which could be macro and small cells \cite{SC_Forum}. 
We also consider the cost of energy by taking into account the power consumption, including that for wireless transmission and that for the operation of base stations. 

Based on this unified model, we 
propose a distributed protocol that improves the spectrum efficiency of the system, which one can
apply weighted proportional fairness among the throughputs of clients, and reduces the cost of energy. Our protocol consists of four components. First, each base station needs to make scheduling decisions for its clients. Second, each base station needs to allocate transmission powers on different frequencies by considering the influence on the throughputs of its clients, the interference caused on others, and the cost of energy. Third, each client needs to choose a suitable base station to be associated with. Finally, each base station needs to determine whether to be in active mode and serve clients, or to be in sleep mode to improve energy efficiency. We propose an online scheduling policy for the first component and shows that it achieves globally optimum performance when the solutions to the other three components are fixed. We also propose distributed strategies for the other three components and show that each of them achieves locally optimal performance under some mild approximation of the system. We show that these strategies only require small computational and communicational overheads, and hence are easily implementable. Moreover, these strategies take the interactions of different components into account. Thus, an integrated solution that applies all these strategies jointly consider all factors of heterogeneous LTE systems.

We also conduct extensive simulations. Simulation results verify that each of the proposed strategies improves system performance. They also show that the integrated solution achieve much better performance than existing policies in a large scale heterogeneous network.

The rest of the paper is organized as follows. Section \ref{section:related} summarizes existing work. Section~\ref{section:system_model} describes the system model and problem setup. Section~\ref{section:scheduling} presents the online scheduling policy for the first component.
Section~\ref{section:power} introduces a distributed heuristic for the second component.
Section~\ref{section:client} discusses both the third and the fourth components, as they are tightly related.
Section~\ref{section:simulation} shows the simulation results.
Finally, Section~\ref{section:conclusion} concludes the paper.

\section{Related Work}	\label{section:related}

There has been some work on self-organized wireless systems. Chen and Baccelli \cite{CSChen10} has proposed a distributed algorithm for the self optimization of radio resources that aims to achieve potential delay fairness. Hu et al \cite{HH10} has proposed a distributed protocol for load balancing among base stations. Borst, Markakis, and Saniee \cite{Sem11} studies the problem of utility maximization for self-organizing networks for arbitrary utility functions. Lopez-Perez et al \cite{DLP11}, Hou and Gupta \cite{Hou11}, and Hou and Chen \cite{Hou12} have considered the problems of jointly optimizing different components in self-organizing networks under various system models. These works do not take energy efficient into considerations.

On the other hand, techniques for
improving cellular radio 
energy efficiency have recently attracted much attention. Auer et al \cite{Auer11} has investigated the amount of power consumptions for various types of base stations. Mclaughlin et al \cite{greenBS11} has discussed various techniques for improving energy efficiency. Conte et al \cite{Conte11} has proposed to turn base stations to sleep mode when the network traffic is small to save energy. Son et al \cite{kson11}, Zhou et al \cite{SZ09}, and Gong, Zhou, and Niu \cite{JG12} have proposed various policies of allocating clients so that clients are mostly allocated to a few base stations. As a result, many base stations that do not have any clients can be turned to sleep mode to save energy. However, these studies require the knowledge of traffic of each client, and cannot be applied to scenarios where clients' traffic is elastic. Chen et al \cite{YC11} has studied the trade-off between spectrum efficiency and energy efficiency. Miao et al \cite{GM09} and Li et al \cite{survey11} have provided extensive surveys on energy-efficient wireless communications. However, they do not consider the interference and interactions between base stations, and are hence not applicable to self-organizing networks.

\section{System Model and Problem Setup}  \label{section:system_model}

Consider a reuse-1 radio system with several base stations and clients that operate and use LTE OFDMA. 
The base stations can be of different types, including macro, micro, pico, and femto base stations. LTE divides frequency bandwidth into subcarriers, and time into frames, which are further divided into 20 time slots. The bandwidth of a subcarrier is 15~kHz while the duration of a time frame is 10~ms. In this paper, we consider LTE frequency division duplex (FDD), the downlink transmission, and resource scheduling. In LTE, a resource block consists of 12 consecutive subcarriers and one time slot of duration 0.5~ms. Under the OFDMA, each user can be allocated any number of resource blocks. However, for each base station, a resource block cannot be allocated to more than one user. LTE can thus achieve both time-division multiplexing and frequency-division multiplexing.

We hereby define $\mathbb{M}$ to be the set of base stations, $\mathbb{I}$ to be the set of clients, and $\mathbb{Z}=\mathbb{F}\times \mathbb{Q}$ to be set of resource blocks, where each $f\in\mathbb{F}$ represents a collection of 12 consecutive subcarriers and each $q\in\mathbb{Q}$ represents a time slot. In the sequel, we use both $z\in\mathbb{Z}$ and $(f,q)$ to denote a resource block for the notational convenience. Note that here we consider reuse-1 systems. However, the result could be extended to other systems.

We consider the energy consumptions of base stations by breaking them into two categories: \emph{operation power} and \emph{transmission power}. When a base station has no clients to serve, the base station can be turned into sleep mode to save power. On the other hand, when the base station has some clients associated to it, it needs to remain in active mode. In addition to transmission power, a base station in active mode also consumes more power for computation, cooling, etc, than one in sleep mode. We call the sum of energy consumption other than transmission power as the \emph{operation power}. We denote by $C_m$ as the difference of operation powers consumed when base station $m\in\mathbb{M}$ is in active mode and when it is in sleep mode.

We denote by $P_{m,z}$ the amount of transmission power that a base station $m\in\mathbb{M}$ assigns on resource block $z$. If base station $m$ does not operate in resource block $z$, we have $P_{m,z}=0$. The time-average transmission power consumed base station $m$ can then be expressed as $\sum_{z\in\mathbb{Z}}P_{m,z}/|\mathbb{Q}|$. Further, we assume that each base station $m$ has a fixed power budget $W_m$ for every time slot, and it is required that $\sum_fP_{m,(f,q)}\leq W_m$, for all $m$ and $q$, which is also known as the per base station transmit power constraint. We note that the values of $C_m$ and $W_m$ can be different from base station to base station, as different types of base stations may consume different amounts of operation powers and have different power budgets. For example, a macro base station has a much larger $C_m$ and $W_m$ than a femto base station.

Propagation loss and path condition are captured by the channel gain. Note that in each resource block, one can consider that the channel gain is usually flat over the subcarriers given that the channel coherence bandwidth is greater than 180 kHz \cite[Ch.12]{LTEbook2011}. It is also time invariant in each time slot given that the channel coherence time is greater than 0.5 ms \cite[Ch.23]{LTEbook2011}. However, the channel gain of a user may change from one resource block to another in the frequency and time domain. Let $G_{i,m,z}$ be the channel gain between base station $m$ and client $i$ on resource block $z$. To be more specific, when the base station $m$ transmits with power $P_{m,z}$, the received power at client $i$ on resource block $z$ is $G_{i,m,z}P_{m,z}$. The received power, $G_{i,m,z}P_{m,z}$, of client $i$ is considered to be its received signal strength if base station $m$ is transmitting data to client $i$, and is considered to be interference, otherwise. Therefore, when base station $m$ is transmitting data to client $i$ on resource block $z$, the signal-to-interference-plus-noise ratio (SINR) of client $i$ on $z$ is expressible as $
\textrm{SINR}_{i,z} =
\frac{G_{i,m,z}P_{m,z}}{N_{i,z}+\sum_{l\neq
m}G_{i,l,z}P_{l,z}},
$
where $N_{i,z}$ is the thermal noise experienced by client $i$ on resource block $z$. The throughput of this transmission can then be described by the Shannon capacity as $B\log(1+\textrm{SINR}_{i,z})$, where $B$ is the bandwidth of a resource block.

Each client $i$ is associated with one base station $m(i)\in\mathbb{M}$. In each frame, base station $m$ schedules one client that is associated with $m$ in each of the resource blocks in the frame. The base station may change the client scheduled in a particular resource block from frame to frame. Let $\phi_{i,m,z}$ be the proportion of frames that client $i$ is scheduled in resource block $z$ by base station $m$. To simplify problem formulation, we assume that $G_{i,m,z}$ does not vary over time. We will then discuss in the following sections how to take channel time variation into account. The influence of channel fading is also demonstrated by simulations in Section \ref{section:simulation}.

Consider that $G_{i,m,z}$ does not vary over the time, the overall throughput of client $i$, which is the sum of its throughput over all the resource blocks, can hence be written as:
\begin{equation}
\mbox{$r_i:=\sum_{z\in\mathbb{Z}}\phi_{i,m(i),z}B\log(1+\frac{G_{i,m,z}P_{m,z}}{N_{i,z}+\sum_{l\neq m}G_{i,l,z}P_{l,z}}).$}
\end{equation}

In this work, we aim to jointly achieve both spectrum efficiency and energy efficiency by considering the tradeoff between them. For spectrum efficiency, we aim to achieve weighted proportional fairness among all the clients when the cost of total power consumption is fixed. Let $w_i$ be the priority weight of client $i$ or user-dependent priority indicator. The weighted proportional fairness can be achieved by maximizing $\sum_{i\in\mathbb{I}}w_i\log r_i$. On the other hand, we also aim to minimize the cost of total power consumption. We denote by $\zeta_m$ as the price of energy for base station $m$. We then formulate the problem of joint spectrum and energy efficiency as the following optimization problem:
\begin{align}
\mbox{Max} & ~~~\mbox{$\sum_{i\in\mathbb{I}}w_i\log r_i -\sum_{m\in\mathbb{M}, z\in\mathbb{Z}}\zeta_m P_{m,z}/|\mathbb{Q}|$}\nonumber\\
&\mbox{$-\sum_{m\in\mathbb{M}: \exists i\in\mathbb{I}\mbox{ s.t. }m(i)=m}\zeta_m C_m$}
\label{equation:introduction:c0}\\
\mbox{s.t.} &~~~\mbox{$\sum_{i:m(i)=m}\phi_{i,m(i),z}= 1, ~~\forall m\in\mathbb{M},z\in\mathbb{Z},$} \label{equation:introduction:c1}\\
&~~~\mbox{$\sum_{f\in\mathbb{F}}P_{m,(f,q)}\leq W_m,$} ~~\forall m\in\mathbb{M},q\in\mathbb{Q}, \label{equation:introduction:c2}\\
\mbox{over}&~~~m(i)\in\mathbb{M}, ~~\forall i\in\mathbb{I},\label{equation:introduction:c3}\\
&~~~P_{m,z}\geq 0, \phi_{i,m(i),z}\geq 0, ~~\forall
i,m,z.\label{equation:introduction:c4}
\end{align}

There are three terms involved in the objective function (\ref{equation:introduction:c0}). The first term, $\sum_{i\in\mathbb{I}}w_i\log r_i$ can be called as the \emph{Weighted Proportional Fairness Index}, as the system achieves weighted proportional fairness by maximizing it. The second term, $\sum_{m\in\mathbb{M}, z\in\mathbb{Z}}\zeta_m P_{m,z}/|\mathbb{Q}|$, is the cost of power consumption on transmission powers of all base stations. In the last term, we note that a base station is only active when it has at least one client, hence, $\sum_{m\in\mathbb{M}: \exists i\in\mathbb{I}\mbox{ s.t. }m(i)=m}\zeta_m C_m$, is the cost of power consumption on operation powers of all base stations. In sum, we aim to maximize $\{$Weighted Proportional Fairness Index$\}-\{$Total cost of power consumption$\}$. In particular, we note that if any of the clients are not covered, i.e., $r_i=0$, for some $i$, then the value of (\ref{equation:introduction:c0}) is $-\infty$. Therefore, by aiming at maximizing (\ref{equation:introduction:c0}), we also guarantee that all clients are covered.

There are two constraints in the formulation. (\ref{equation:introduction:c1}) states that, for each base station, it can only allocate a resource block to one client in each frame. However, for a fixed resource block, the base station may change the client that it is allocated to from frame to frame. The second constraint, (\ref{equation:introduction:c2}), states that the total amount of power that a base station allocates on all subcarriers cannot exceed its power budget. The variables that we are able to control are listed in (\ref{equation:introduction:c3}) and (\ref{equation:introduction:c4}), which include the base station that each client is associated to, $m(i)$, the transmission power that each base station allocates on each resource block, $P_{m,z}$, and the scheduling decision of each base station on each resource block, $\phi_{i,m(i),z}$. Finally, we note that a base station only needs to be in active mode when at least one client is associated with it, and can be in sleep mode when no clients are associated with it. Therefore, the decision on whether a base station is in sleep mode or in active mode is implicitly determined by the choices of $m(i)$, for all $i$.

This formulation shows that there are several important components involved. In each frame, a base station needs to decide which client should be scheduled in each resource block. This essentially determines the values of $\phi_{i,m(i),z}$ so as to maximize $\sum_{i\in\mathbb{I}}w_i\log r_i$. We call this component the \emph{Scheduling Problem}. In each frame, a base station also needs to decide how much power it should allocate in each resource block, subject to the constraint on its power budget. This component is referred as the \emph{Power Control Problem}. The Power Control Problem influences both the spectrum efficiency, $\sum_{i\in\mathbb{I}}w_i\log r_i$, and the cost of transmission power, $\sum_{m\in\mathbb{M}, z\in\mathbb{Z}}\zeta_m P_{m,z}$. Besides, every client needs to choose an active base station to be associated with. Base stations also need to decide whether to be in active mode to serve clients, or to be in sleep mode to save energy. We denote both the clients' decisions on associated base stations and base stations' decisions on whether to be in active or sleep mode as the \emph{Client Association Problem}, as these two components are tightly related. Hence, the Client Association Problem influences both spectrum efficiency and the total cost of operation power, $\sum_{m\in\mathbb{M}: \exists i\in\mathbb{I}\mbox{ s.t. }m(i)=m}\zeta_m C_m$.

\begin{figure}[t]
\centering
\includegraphics[width=0.3\textwidth]{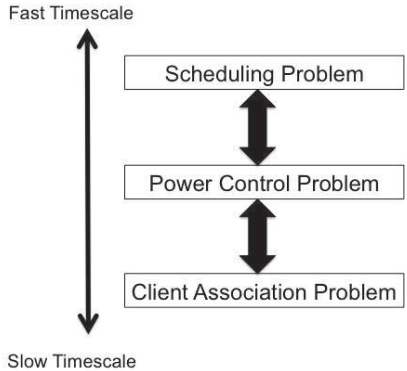}
\caption{Solution overview.}
\label{fig:overview}
\end{figure}

Further, we notice that there is a natural timescale separation between the three components: The Scheduling Problem is updated on a per time slot basis. On the other hand, the Power Control Problem is updated in a slower timescale. Finally, the Client Association Problem must only be updated infrequently, as the overheads for clients to change the associated base stations, and for the base stations to switch between sleep/active mode are large. In the following, we first propose an online algorithm for the Scheduling Problem, given solutions to the Power Control Problem and the Client Association Problem. We then propose a heuristic for the Power Control Problem by considering solutions to the Scheduling Problem. Finally, we develop a protocol for the Client Association Problem. The protocol uses the knowledge of the Power Control Problem as well as the influences on the Scheduling Problem. Figure \ref{fig:overview} illustrates an overview of our approach and the timescales of different components.

\section{Online Algorithm for the Scheduling Problem}
\label{section:scheduling}

In this section, we study the Scheduling Problem, given solutions to the Power Control Problem and the Client Association Problem, i.e., values of $P_{m,z}$ and $m(i)$. We thus define
\begin{equation}    \label{equation:scheduling:H}
H_{i,m(i),z}:=B\log(1+\frac{G_{i,m(i),z}P_{m(i),z}}{N_{i,z}+\sum_{l\neq m(i)}G_{i,l,z}P_{l,z}}),
\end{equation}
which is the throughput of client $i$ on resource block $z$ when it is scheduled by base station $m(i)$, to simplify the notations. With $P_{m,z}$ and $m(i)$ fixed, solving (\ref{equation:introduction:c0})-(\ref{equation:introduction:c4}) is equivalent to solving the following:
\begin{align}
\mbox{Max} &~~\mbox{$\sum_{i\in\mathbb{I}} w_i\log r_i=\sum_{i\in\mathbb{I}} w_i\log(\sum_{z\in\mathbb{Z}}\phi_{i,m(i),z}H_{i,m(i),z})$}\label{equation:scheduling:c0}\\
\mbox{s.t.} &~~\mbox{$\sum_{i:m(i)=m}\phi_{i,m(i),z}= 1, \forall m\in\mathbb{M},z\in\mathbb{Z},$} \label{equation:scheduling:c1}\\
&~~\phi_{i,m(i),z}\geq 0, \forall i,z.\label{equation:scheduling:c2}
\end{align}

One can see that the above optimization problem is in fact convex and hence can be solved by standard techniques of convex optimization. To further simplify the computation overhead, we propose an online scheduling policy for the Scheduling
Problem. Let $\phi_{i,m(i),z}[k]$ be the proportion of frames that base station $m(i)$ has scheduled client $i$ for resource block $z$ in the first $k-1$ frames. Similarly, let $r_i[k]=\sum_{z\in\mathbb{Z}}\phi_{i,m(i),z}[k]H_{i,m(i),z}$ be the average throughput of client $i$ in the first $k$ frames. We then have $\phi_{i,m(i),z}[k+1]  = \frac{k-1}{k}\phi_{i,m(i),z}[k]+\frac{1}{k}$, if client $i$ is scheduled for $z$ in the $(k+1)$-th frame, and $\phi_{i,m(i),z}[k+1]  =  \frac{k-1}{k}\phi_{i,m(i),z}[k],$ otherwise.


In our online scheduling policy, the base station schedules the client $i$ that maximizes $w_iH_{i,m(i),z}/r_i[k]$. This is indeed the well-known proportional fair scheduling \cite{HK04}, and it has been shown that, when the solutions to the Power Control Problem and the Client Association Problem are fixed, the online scheduling policy converges to the optimal solution to the Scheduling Problem.

\begin{theorem}[\cite{HK04}] \label{theorem:scheduling:optimal}
Using the above scheduling policy, the value of
$\liminf_{k\rightarrow \infty}\sum_{i\in\mathbb{I}}w_i\log r_i[k]$
achieves the maximum of the optimization problem
(\ref{equation:scheduling:c0})--(\ref{equation:scheduling:c2}).
\end{theorem}


Note that in the previous discussions, we have assumed that the channel gain, $G_{i,m,z}$, does not vary over time. In practice, however, channel gains fluctuate due to fading. To take fading into account, we let $\hat{G}_{i,m,z}$ be the instantaneous time-varying channel gain, and $\hat{H}_{i,m(i),z}$ be the instantaneous throughput that client $i$ can get from resource block $z$ if it is scheduled by base station $m(i)$. Our scheduling policy can then be easily modified such that the base station schedules the client with the largest $w_i\hat{H}_{i,m(i),z}/r_i[k]$ on resource block $z$. In Section \ref{section:simulation}, we show that this modification can further improve performance.

\section{A Heuristic for the Power Control Problem}
\label{section:power}

In this section, we discuss the Power Control Problem, i.e., how the base stations choose $P_{m,z}$ so as to solve (\ref{equation:introduction:c0})--(\ref{equation:introduction:c4}).

Obviously, base stations need to know the solution of the Scheduling Problem
$\{\phi_{i,m,z}\}$ and the values of channel gains
$\{G_{i,m,z}\}$, in order to choose suitable $P_{m,z}$. To
reduce computation and communication overhead and maintain operational simplicity, the base stations assume that
for all clients $i$ associated with base station $m$, the perceived thermal
noises are all $N_{m,z}$, the channel gains between them and base station $m$
are all $G_{m,m,z}$, and the channel gains between them and base station $l~(\neq m)$
are all $G_{m,l,z}$ on resource block $z$. In practice, $N_{m,z}$ and
$G_{m,m,z}$ can be set to be the average value of the noise powers and the average value of the channel gains
between $m$ and its clients, respectively, and
$G_{m,l,z}$ can be set to be the
channel gain between base station $m$ and base station $l$ on resource block $z$. Under this
assumption, one can see that the optimal solution to the Scheduling Problem (\ref{equation:scheduling:c0})--(\ref{equation:scheduling:c2}) is given by:
$\phi_{i,m(i),z}= \frac{w_i}{\sum_{j:m(j)=m(i)}w_j}.$

Let $w^m:=\sum_{i:m(i)=m}w_i$ be the sum weight of the clients that are associated with base station $m$. Under a fixed solution to the Client Association Problem and the above approximation to the Scheduling Problem, (\ref{equation:introduction:c0})-(\ref{equation:introduction:c4}) can be rewritten as:
\begin{align}
\mbox{Max} &~\mbox{$\sum_{m\in\mathbb{M}}w^m\log\sum_{z\in\mathbb{Z}}B\log(1+\frac{G_{m,m,z}P_{m,z}}{N_{m,z}+\sum_{l\neq m}G_{m,l,z} P_{l,z}})$}\nonumber\\
&\mbox{$- \sum_{m\in\mathbb{M},z\in\mathbb{Z}}\zeta_mP_{m,z}/|\mathbb{Q}|$}
\label{equation:power:c0}\\
\mbox{s.t.} &~\mbox{$\sum_{f\in\mathbb{F}}P_{m,(f,q)}\leq W, \forall m\in\mathbb{M},q\in\mathbb{Q},$} \label{equation:power:c1}\\
&~P_{m,z}\geq 0, \forall i,m,z.\label{equation:power:c2}
\end{align}

This problem is non-convex and it may be infeasible to find the global optimal solution. Instead, we propose a distributed heuristic that employs the gradient method, which converges to a local optimum\cite{Bazaraa}. 
Let $P$ be the vector consisting of $\{P_{m,z}\}$, $\textrm{SINR}_{m,z}(P):= \frac{G_{m,m,z}P_{m,z}}{N_{m,z}+\sum_{l\neq m} G_{m,l,z} P_{l,z}},$ and $U(P):=\sum_{m\in\mathbb{M}}w^m\log[\sum_{z\in\mathbb{Z}}B\log(1+\textrm{SINR}_{m,z}(P))]- \sum_{m\in\mathbb{M},z\in\mathbb{Z}}\zeta_mP_{m,z}/|\mathbb{Q}|.$ We have:
\begin{align*}
&\frac{\partial U(P)}{\partial P_{m,z}}\\
=& \frac{w^m}{\sum_{y\in\mathbb{Z}} \log(1+\textrm{SINR}_{m,y}(P))} \times\frac{G_{m,m,z}}{N_{m,z}+\sum_{l} G_{m,l,z} P_{l,z}}\\
&+\sum_{o\neq m}\frac{w^o}{\sum_{y\in\mathbb{Z}}\log(1+\textrm{SINR}_{o,y}(P))}\times[\frac{G_{o,m,z}}{N_{o,z}+\sum_{l}G_{o,l,z}P_{l,z}}\\
&-\frac{G_{o,m,z}}{N_{o,z}+\sum_{l\neq o}G_{o,l,z}P_{l,z}}]-\zeta_m/|\mathbb{Q}|.
\end{align*}

Each base station updates its power periodically. When base station $m$ updates its power, it sets its power on resource block $(f,q)$ to be:
\begin{align}
\left\{
\begin{array}{l}
[P_{m,(f,q)}+\alpha
\frac{\partial U(P)}{\partial P_{m,(f,q)}}]^+ ,\\
\hspace{20pt}~~\textrm{if}~ \sum_e[P_{m,(e,q)}+\alpha \frac{\partial U(P)}{\partial P_{m,(e,q)}}]^+ \leq W_m,\\\\
W\frac{[P_{m,(f,q)}+\alpha\frac{\partial U(P)}{\partial P_{m,(f,q)}}]^+}{\sum_e[P_{m,(e,q)}+\alpha \frac{\partial U(P)}{\partial P_{m,(e,q)}}]^+}, ~~\textrm{otherwise},
\end{array}
\right.
\end{align}
where $x^+:=\max\{x,0\}$ and $\alpha$ is a
small constant. Base station $m$ needs to compute $\frac{\partial
U(P)}{\partial P_{m,z}}$ to update its power on each resource block. The
computation of $\frac{\partial U(P)}{\partial P_{m,z}}$ can be
further simplified
by setting $G_{o,m,z}=0$ for all $o$ such that
$G_{o,m,z}$ is small and has little influence on the value of
$\frac{\partial U(P)}{\partial P_{m,z}}$. Thus, to compute
$\frac{\partial U(P)}{\partial P_{m,z}}$,
base station $m$ exchanges information with base station $o$ that is physically close to it
so as to know:
\begin{itemize}
    \item the sum weight of clients associated with $o$, i.e., $w^o$,
    \item the channel gain $G_{o,m,z}$ from $m$ to $o$,
    \item the sum of interference and noise $N_{o,z}+\sum_{l\neq o}G_{o,l,z}P_{l,z}$ at $o$,
    \item the received signal strength $G_{o,o,z}P_{o,z}$ at $o$, and
    \item the average total throughput $\sum_{y\in\mathbb{Z}}\log(1+\textrm{SINR}_{o,y}(P))$ in the downlink of base station~$o$,
\end{itemize}
for all $o$ that $G_{o,m,z}$ is large. In LTE, the above information
can be obtained through periodic channel quality indicator
and reference signal reports.

This method is easy to implement and only requires limited
information exchange between neighbor cells.
We assume that the neighbor cell
communication takes place between base stations and is supported by
the wired backhaul network.

\section{A Protocol for the Client Association Problem}
\label{section:client}

In the following, we discuss how to solve the Client Association Problem, i.e., how each client $i$ should choose a base station $m(i)$. Our solution consists of two parts: In the first part, each client $i$ estimates its throughput when it associates with each base station. The client $i$ then selfishly chooses the base station that maximizes its throughput. In the second part, each base station decides whether to be in active mode or in sleep mode by jointly considering the effects on spectrum efficiency and energy consumption.

\subsection{A Selfish Strategy of Clients}

We assume that each client $i$ is selfish and would like to choose a base station $m$ that maximizes its own throughput when it is associated with $m$. We make this assumption under two main reasons. First, this conforms to the selfish behaviors of clients. Second, in a dense network, the decision of $m(i)$ by one client $i$ only has a limited and indirect impact on the overall performance of other clients.

We define $H_{i,m,z}$ as in (\ref{equation:scheduling:H}), and $\hat{\phi}_{i,m,z}$ be the proportion of frames that base station $m$ would schedule $i$ if $i$ is associated with $m$. The client $i$ then selects the base station $m$ that maximizes
$
\hat{r}_{i,m} := \sum_{z\in\mathbb{Z}}\hat{\phi}_{i,m,z}H_{i,m,z}.
$

In practice, client $i$ can only be associated with base stations that are in active mode and whose $H_{i,m,z}$ is above some threshold for some $z\in\mathbb{Z}$. To compute $\hat{r}_{i,m}$ for all such base stations, client $i$ needs to know the values of $H_{i,m,z}$ and $\hat{\phi}_{i,m,z}$. Client $i$ assumes that the transmission powers used by base stations are not influenced much by its choice, which is true in a dense network. Thus, client $i$ only needs to know its perceived SINR with each base station on each resource block to compute $H_{i,m,z}$. It remains for the client $i$ to compute the value of $\hat{\phi}_{i,m,z}$. We propose two different approaches to compute this value. In the first approach, which we call the \emph{Exact Simulator} (ES), client $i$ first obtains the values of $w_j$ and $H_{j,m,z}$ for all clients $j$ that are associated with $m$. Client $i$ can then simulates the scheduling decisions of base station $m$ by running the online scheduling policy introduced in Section \ref{section:scheduling}, and obtains the value of $\hat{\phi}_{i,m,z}$ on each resource block $z$. While this approach offers an accurate estimation on $\hat{\phi}_{i,m,z}$ and $\hat{r}_{i,m}$, it requires high computation and communication overhead.

In the second approach, which we call the \emph{Approximate Estimator} (AE), client $i$ only obtains the values of $w^m_{-i}:=\sum_{j:j\neq i, m(j)=m}w_j$, and
$
\overline{H}_{m,z}:=\sum_{j:m(j)=m}\phi_{j,m,z}H_{j,m,z},
$
which is the average throughput of base station $m$ on resource block $z$. Client $i$ assumes that, when another client $j$ is scheduled by base station $m$ on resource block $z$, its throughput on $z$ equals the average throughput $\overline{H}_{m,z}$. Client $i$ can then estimate $r_i$ by \emph{Algorithm~\ref{algorithm:client:AE}}. The complexity of \emph{Algorithm~\ref{algorithm:client:AE}} is $O(|\mathbb{Z}|)$, and therefore this approach is much more efficient than the Exact Simulator. Moreover, the following theorem suggests that the Approximate Simulator provides reasonably good estimates on the throughput of client $i$ if it is associated with base station $m$.

\begin{algorithm}[t]
\caption{Approximate Estimator} \label{algorithm:client:AE}
\begin{algorithmic}[1]
\STATE Sort all resource blocks such that
$\frac{\overline{H}_{m,1}}{H_{i,m,1}}\leq \frac{\overline{H}_{m,2}}{H_{i,m,2}}\leq \frac{\overline{H}_{m,3}}{H_{i,m,3}}\leq\dots$\\
\STATE $\hat{\phi}_{i,m,z}\leftarrow 0, \forall z\in\mathbb{Z}$
\STATE $\hat{r}_{i,m} \leftarrow 0,\forall i\in\mathbb{I}$
\STATE $\overline{r}_m\leftarrow \sum_{z\in\mathbb{Z}} \overline{H}_{m,z}$\\
\FOR{$z=1 \to |\mathbb{Z}|$}

\IF{$\frac{w_iH_{i,m,z}}{\hat{r}_{i,m}+H_{i,m,z}}>\frac{w^m_{-i}\overline{H}_{m,z}}{\overline{r}_m-\overline{H}_{m,z}}$}
\STATE $\hat{\phi}_{i,m,z}\leftarrow 1$\\
\STATE $\hat{r}_{i,m}\leftarrow \hat{r}_{i,m}+H_{i,m,z}$\\
\STATE $\overline{r}_m\leftarrow \overline{r}_m-\overline{H}_{m,z}$
\ELSIF{$\frac{w_iH_{i,m,z}}{\hat{r}_{i,m}}<\frac{w^m_{-i}\overline{H}_{m,z}}{\overline{r}_m}$}
\STATE Break \ELSE
\STATE $\hat{\phi}_{i,m,z}\leftarrow\frac{\overline{r}_mw_iH_{i,m,z}-\hat{r}_{i,m}w^m_{-i}\overline{H}_{m,z}}{(w^m_{-i}+w_i)H_{i,m,z}\overline{H}_{m,z}}$\\
\STATE $\hat{r}_{i,m}\leftarrow \hat{r}_{i,m}+\hat{\phi}_{i,m,z}H_{i,m,z}$\\
\STATE $\overline{r}_m\leftarrow
\overline{r}_m-\hat{\phi}_{i,m,z}\overline{H}_{m,z}$ \STATE Break \ENDIF
\ENDFOR
\RETURN $\hat{r}_{i,m}$
\end{algorithmic}
\end{algorithm}

\begin{theorem} \label{theorem:client:AE}
If for each client $j$ other than $i$ that is associated $m$,
$H_{j,m,z}=\overline{H}_{m,z}$, then, under the online scheduling
policy introduced in Section \ref{section:scheduling}, the
throughput of client $i$ equals the value of $\hat{r}_{i,m}$ obtained by
Algorithm \ref{algorithm:client:AE} when it is also associated with
$m$.
\end{theorem}

\begin{IEEEproof}
Please refer to Appendix B for the proof.
\end{IEEEproof}

After obtaining the value of $\hat{r}_{i,m}$, client $i$ selects $m(i)=\arg\max_m \hat{r}_{i,m}$ and associates with base station $m(i)$. Moreover, client $i$ reports the estimated rate with $m(i)$, $\hat{r}_{i,m(i)}$, and the second largest estimated rate among all base stations, $\max_{m\neq m(i)}\hat{r}_{i,m}$, to $m(i)$. We define $m^*(i):=\arg\max_{m\neq m(i)}\hat{r}_{i,m}$, and hence $\hat{r}_{i,m^*(i)}=\max_{m\neq m(i)}\hat{r}_{i,m}$. These values are used for base stations to decide whether to switch to sleep mode, which is discussed in the following section.

In summary, when the Approximate Estimator is applied, each base station $m$ only needs to periodically broadcasts a total number of $|\mathbb{Z}|+1$ values, that is, the values of $w^m$, from which each client $i$ can compute $w^m_{-i}$, and $\overline{H}_{m,z}$ for all $z\in\mathbb{Z}$. On the other hand, when a client $i$ decides to be associated with a base station $m(i)$, it only needs to report two values: $\hat{r}_{i,m(i)}$ and $\hat{r}_{i,m^*(i)}$. Thus, the communication overhead for the Approximate Estimator is small.

\subsection{A Distributed Protocol of Base Stations}

We now discuss how base stations decide whether to be in active mode or in sleep mode. The protocol consists of two parts: one for a base station in active mode to decide whether to switch to sleep mode, and one for a base station in sleep mode to decide whether to switch to active mode.

First, consider a base station $m_0$ that is in active mode. Given the solution to the Power Control Problem, $m_0$ aims to maximize $\sum_{i:m(i)=m_0}w_i\log r_i-\zeta_{m_0}C_{m_o}1\{\mbox{$m_0$ is in active mode}\}$, where $1\{\cdot\}$ is the indicator function. When $m_0$ is in active mode, it estimates that $r_i=\hat{r}_{i,m(i)}$, which is the estimated throughput that client $i$ reports to $m_0$ as discussed in the previous section. Therefore, we have:
\begin{align}	
&\sum_{i:m(i)=m_0}w_i\log r_i-\zeta_{m_0}C_{m_o}1\{\mbox{$m_0$ is in active mode}\}\nonumber\\
=&\sum_{i:m(i)=m_0}w_i\log \hat{r}_{i,m(i)}-\zeta_{m_0}C_{m_o}.\label{equation:client:active to sleep 1}
\end{align}
On the other hand, when $m_0$ is in sleep mode, it assumes that its clients will be associated with the base station other than $m_0$ that provides the largest throughput, and the resulting throughput of client $i$ will be $\hat{r}_{i,m^*(i)}$. We then have:
\begin{align}
&\sum_{i:m(i)=m_0}w_i\log r_i-\zeta_{m_0}C_{m_o}1\{\mbox{$m_0$ is in active mode}\}\nonumber\\
=&\sum_{i:m(i)=m_0}w_i\log \hat{r}_{i,m^*(i)}.\label{equation:client:active to sleep 2}
\end{align}
The base station $m_0$ simply compares the values of (\ref{equation:client:active to sleep 1}) and of (\ref{equation:client:active to sleep 2}). If the latter is larger than the former, $m_0$ switches to sleep mode. 

Next, consider a base station $m_1$ that is in sleep mode. Base station $m_1$ periodically wakes up and broadcasts beacon messages on all resource blocks, where it uses equal amounts of power on all resource blocks. Each client $i$ then measures the SINR from $m_1$ on each resource block and obtains the values of $H_{i,m_1,z}$ for all $z$. Each client $i$ computes $\hat{r}_{i,m_1}=\sum_{z\in\mathbb{Z}}H_{i,m_1,z}$, which is the estimated throughput of $i$ if $i$ is the only client associated with $m_1$. If $\hat{r}_{i,m_1}$ is larger than $r_i$, the current throughput of client $i$, client $i$ reports the values of $w_i, \hat{r}_{i,m_1}$, and $r_i$ to base station $m_1$.

Base station $m_1$ needs to estimate the throughput of clients that will be associated with it when it switches to active mode. We note that the value of $\hat{r}_{i,m_1}$ is not a good estimate of the throughput of $i$ when $m_1$ is in active mode and $i$ associates with $m_1$. Recall that $\hat{r}_{i,m_1}$ is the estimated throughput of client $i$ when $i$ is the only client associated with $m_1$. When $m_1$ is in active mode, it is possible that $m_1$ will have more than one clients, and hence the throughput of $i$ may be much less than $\hat{r}_{i,m_1}$ when $i$ is associated with $m_1$. To better estimate clients' throughputs when $m_1$ is in active mode, $m_1$ assumes that frequency-selective fading is not significant, and the values of $H_{i,m_1,z}$ is the same for all $z$ when $i$ is fixed. Under this assumption, our online scheduling policy will result in $\phi_{i,m_1,z}=\frac{w_i}{w^{m_1}}$ for all clients that are associated with $m_1$ and $z\in\mathbb{Z}$, where $w^{m_1}=\sum_{i:m(i)=m_1}w_i$. Base station $m_1$ then runs Algorithm \ref{algorithm:client:basestation} and estimates that the set $S$ of clients that will be associated with $m_1$ when $m_1$ switches to active mode. Hence $w^{m_1}$ is estimated to be $\hat{w}^{m_1}$, and each client $i$ in $S$ is estimated to have a throughput of $\hat{r}_{i,m_1}\phi_{i,m_1,z}=\hat{r}_{i,m_1}\frac{w_i}{w^{m_1}}$.

\begin{algorithm}[t]
\caption{Wakeup Estimator} \label{algorithm:client:basestation}
\begin{algorithmic}[1]
\STATE Sort all clients that report to $m_1$ such that $\frac{\hat{r}_{1,m_1}}{r_1}w_1\geq\frac{\hat{r}_{2,m_1}}{r_2}w_2\geq\dots\geq \frac{\hat{r}_{j,m_1}}{r_j}w_j$
\STATE $S\leftarrow$ NULL
\STATE $\hat{w}^{m_1}\leftarrow 0$
\FOR{$i=1\to j$}
\IF{$\frac{\hat{r}_{i,m_1}}{r_i}w_i > \hat{w}^{m_1}+w_i$}
\STATE $S\leftarrow S\cup\{i\}$
\STATE $\hat{w}^{m_1}\leftarrow \hat{w}^{m_1}+w_i$
\ENDIF
\ENDFOR
\RETURN $S, \hat{w}^{m_1}$
\end{algorithmic}
\end{algorithm}

We now discuss the intuitions of Algorithm \ref{algorithm:client:basestation}. Let $i^*$ be the largest value in $S$, and thus the value of $\hat{w}^{m_1}$ is last updated in Line 7 of the $i^*$-th iteration in Algorithm \ref{algorithm:client:basestation}. We then have $\frac{\hat{r}_{i^*,m_1}}{r_{i^*}}w_{i^*} > \hat{w}^{m_1}=w^{m_1}$. For each $i$ in $S$, we have $\frac{\hat{r}_{i,m_1}}{r_{i}}w_{i} \geq\frac{\hat{r}_{i^*,m_1}}{r_{i^*}}w_{i^*} > w^{m_1}$, and hence $\hat{r}_{i,m_1}\frac{w_i}{w^{m_1}}>r_i$, that is, client $i$ is estimated to have higher throughput when it is associated with $m_i$. This justifies the estimation that client $i$ will leave its current base station to be associated with $m_1$. On the other hand, for each client $i'$ that is not in $S$, we have $\frac{\hat{r}_{i',m_1}}{r_{i'}}w_{i'}\leq w^{m_1}$ and $\hat{r}_{i',m_1}\frac{w_{i'}}{w^{m_1}}\leq r_{i'}$, justifying the estimation that $i'$ will stay with its current base station and not be associated with $m_1$. Therefore, Algorithm \ref{algorithm:client:basestation} provides a reasonable estimation on the set of clients that will be associated with $m_1$ when $m_1$ switches to active mode. Moreover, the complexity of Algorithm \ref{algorithm:client:basestation} is only $O(j\log j)$, where $j$ is the number of clients that report to $m_1$.

Finally, base station $m_1$ compares the value of $\sum_{i\in S}w_i\log(\hat{r}_{i,m_1}\frac{w_i}{w^{m_1}})-\zeta_{m_1}C_{m_1}$ against the value of $\sum_{i\in S}w_i\log r_i$. If the former is larger, that is,
$
\sum_{i\in S}w_i[\log(\hat{r}_{i,m_1}\frac{w_i}{w^{m_1}})-\log r_i]-\zeta_{m_1}C_{m_1}>0,
$
base station $m_1$ switches to active mode.

\section{Simulation Results}    \label{section:simulation}

In the following, we present the simulation results. We first present simulation results on three simple and small-scale systems. While these systems may be over-simplified, they provide insights on our solutions for the components discussed in the paper, namely, the Scheduling Problem, the Power Control Problem, and the Client Association Problem. We then present simulation results for a large-scale system.

We consider a reuse-1 LTE-FDD system with a bandwidth of 10 MHz, which can accommodate 600 subcarriers \cite{3GPP2010R10UE} and there are $\frac{600}{12} \times 20 = 1000$ resource blocks. Channel gains are derived from the following equation:
\begin{equation}
PL(d) = 128.1 + 37.6 \log_{10}(d) + X + Y,
\end{equation}
where $PL(d)$ is the channel gain in dB and $d$ is distance in km. $X$ and $Y$ represent shadowing and fast fading, respectively. $X$ is the log-normal shadowing with mean 0 and standard deviation 8 dB. Since $X$ is a slow fading, we consider that it is time invariant. However, it will vary in frequency, in every 180 kHz. On the other hand, $Y$ represents Rayleigh fast fading with a Doppler of 5 Hz. It also varies in frequency. The thermal noise is randomly generated in $[3.5, 4.5]\times 10^{-15}$~W.

Auer et al. \cite{Auer11} has investigated the amount of power needed to operate a base station. It has shown that a macro base station consumes 75W when it is in sleep mode, and consumes 130W when it is in active mode. Therefore, we set the operation power of a macro base station to be 55W. It has also shown that the power budget of a macro base station is 20W. Similarly, we set the operation power and transmission power budget of a micro base station to be 17W and 6.3W, respectively.

We have implemented the proposed online scheduling policy, power control, and Approximate Estimator. We compare our mechanisms against other mechanisms. We consider two policies for the Scheduling Problem, round-robin (\emph{RR}) and the scheduling policy proposed in Section \ref{section:scheduling} (\emph{PF}). For the Power Control Problem, we assume that other mechanisms use the same amount of power on each resource block. We consider two policies for the Client Association Problem. One of the policies associate each client to the closest base station, and is called \emph{Default}. The other policy adapts the ones proposed in Son et al \cite{kson11} and Zhou et al \cite{SZ09}. In this policy, which we call \emph{Son-Zhou}, each client $i$ chooses to be associated with the base station $m$ that maximizes \{Data rate of $i$ when served by $m$\}$\times$\{Number of clients associated $m$\}/\{Operation power of $m$\}. The intuition is that clients prefer to be associated with base stations with many clients. As a result, some of the base stations will have very few clients and can be turned into sleep mode. We assume that a base station is turned into sleep mode if the total weight of its clients is below a certain threshold. We have exhaustively evaluated the performance of Son-Zhou using different thresholds and found that setting the threshold to be 2 achieves the best performance under all evaluated price of energy. Hence, we set the threshold to be 2.

We compare the performance of different mechanisms by their resulting values of 
(\ref{equation:introduction:c0}), where the throughput of a client, $r_i$, is measured in kbits/sec.  We also evaluate and present the achieved total weighted throughput, defined as $\sum_i w_i r_i$, total power consumption, and/or energy efficiency of each mechanism under various scenarios.

\begin{figure}[t]
\centering
\includegraphics[width=0.45\textwidth]{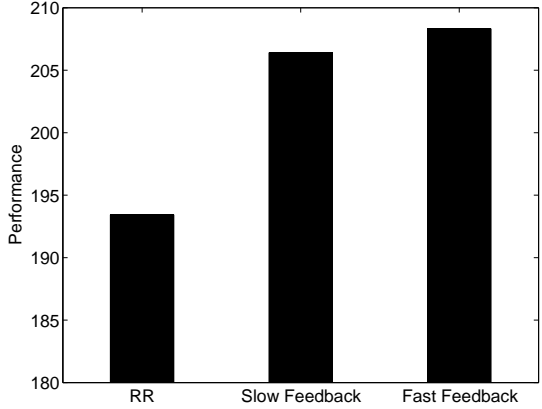}
\caption{Performance comparison for scheduling.}
\label{fig:scheduling_performance}
\end{figure}

\begin{figure}[t]
\centering
\includegraphics[width=0.45\textwidth]{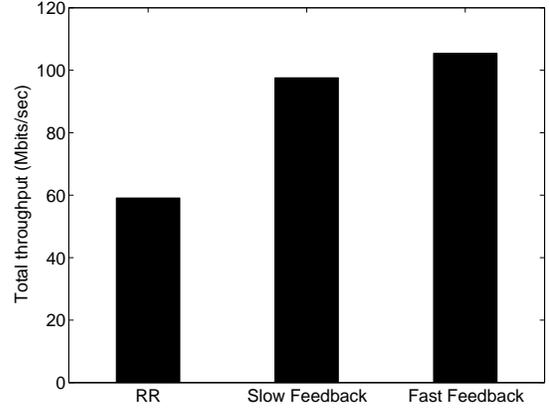}
\caption{Throughput comparison for scheduling.}
\label{fig:scheduling_throughput}
\end{figure}

We first consider a system with one macro base station and 25 clients to demonstrate the performance of our solution to the Scheduling Problem. Clients are uniformly placed as a $5\times 5$ grid, where the distance between adjacent clients is 100m. We compare the RR policy against our PF policy, where we consider both cases when the base station has instant knowledge of channel gains and where base stations only have knowledge of long-term average channel gains, denoted by \emph{Fast Feedback} and \emph{Slow Feedback}, respectively. We set the price of energy, $\zeta_m$, to be zero for this system, as we are only interested in the performance of {scheduling} policies. Figure \ref{fig:scheduling_performance} and Figure \ref{fig:scheduling_throughput} show the simulation results on both the values of (\ref{equation:introduction:c0}) and the total throughput of the system. It 
is observed that both Fast Feedback and Slow Feedback achieve more than 50\% higher throughput than the RR policy, as they both use knowledge on channel gains for scheduling decisions. Fast Feedback has better performance than Slow Feedback since it takes effects of fast fading into account.

\begin{figure}[t]
\centering
\includegraphics[width=0.45\textwidth]{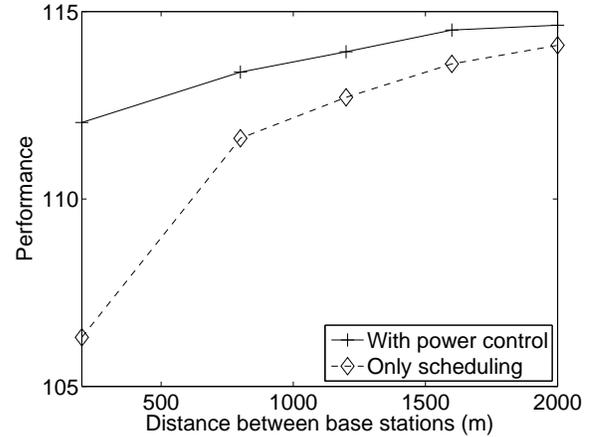}
\caption{Performance comparison for power control.}
\label{fig:power_performance}
\end{figure}

\begin{figure}[t]
\centering
\includegraphics[width=0.45\textwidth]{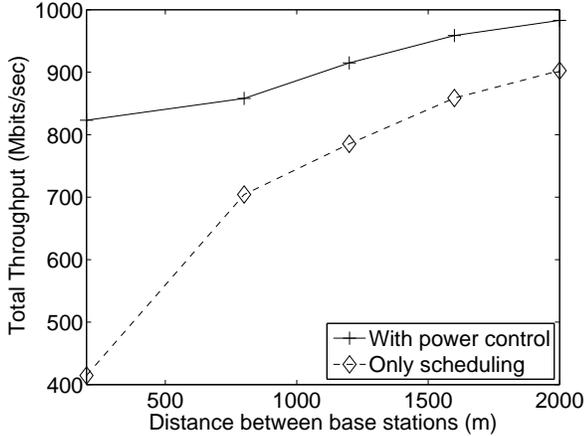}
\caption{Throughput comparison for power control.}
\label{fig:power_throughput}
\end{figure}

\begin{figure}[t]
\centering
\includegraphics[width=0.45\textwidth]{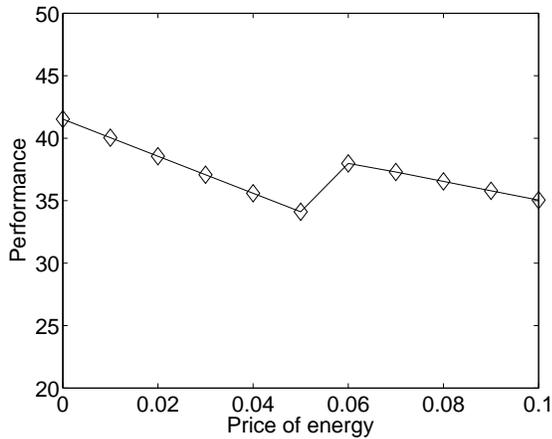}
\caption{System performance under various energy prices.}
\label{fig:association_performance}
\end{figure}

\begin{figure}[t]
\centering
\includegraphics[width=0.45\textwidth]{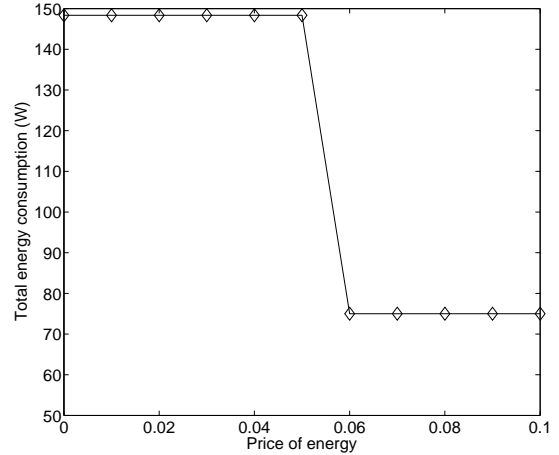}
\caption{Power consumption under various energy prices.}
\label{fig:association_consumption}
\end{figure}

\begin{figure}[t]
\centering
\includegraphics[width=0.45\textwidth]{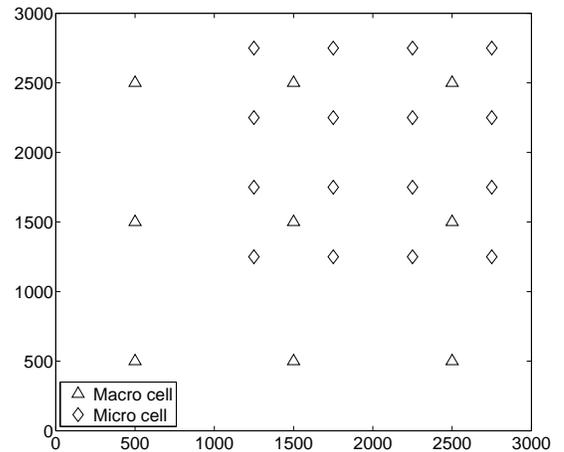}
\caption{Topology of the simulation.}
\label{fig:grid}
\end{figure}

\begin{figure}[t]
\centering
\includegraphics[width=0.45\textwidth]{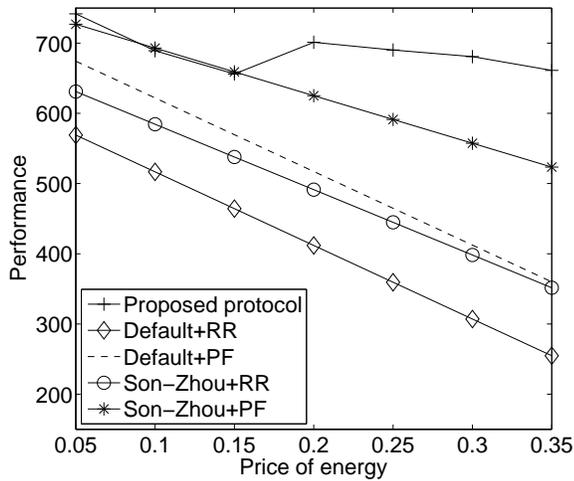}
\caption{Performance comparison under various energy prices.}
\label{fig:full}
\end{figure}

\begin{figure}[t]
\centering
\includegraphics[width=0.45\textwidth]{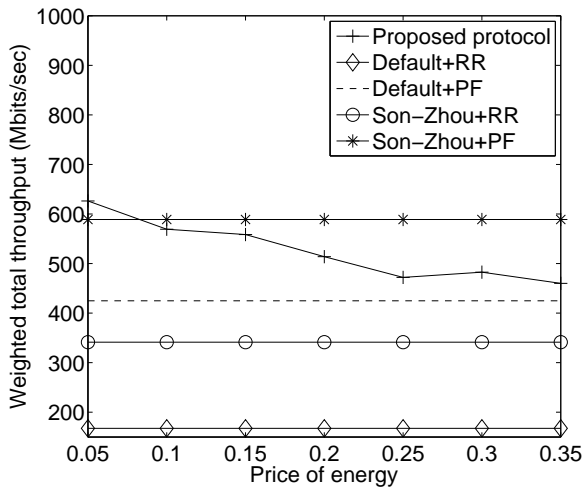}
\caption{Total weighted throughput under various energy prices.}
\label{fig:spectrum_efficiency}
\end{figure}

\begin{figure}[t]
\centering
\includegraphics[width=0.45\textwidth]{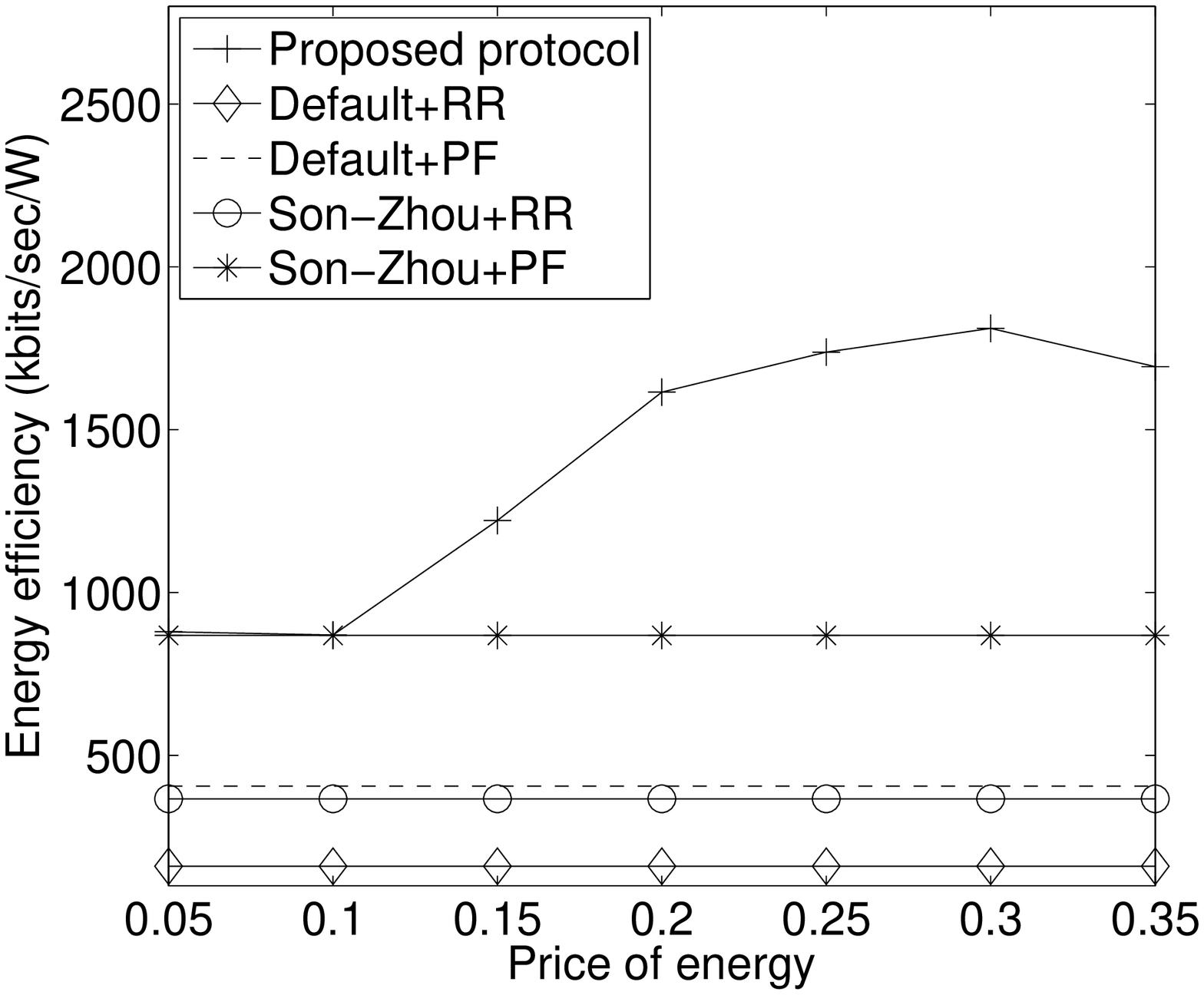}
\caption{Energy efficiency under various energy prices.}
\label{fig:energy_efficiency}
\end{figure}

Next, we demonstrate the performance of our solution to the Power Control Problem. We consider a system with two macro base stations. Each of these base stations has two clients associated with it, and the distance between a client and its associated base station is 50m. We compare a policy that uses both our scheduling policy and power control algorithm against one that only uses our scheduling policy and allocates equal power on all resource blocks. We consider the performance of the two policies by varying the distance between the two base stations. We also set the price of energy to be zero for this system. Simulation results are shown in Figure 
\ref{fig:power_performance} and Figure \ref{fig:power_throughput}. It 
is observed that when the two base stations are far apart, the two policies achieve similar performance. However, as the distance between the two base stations decreases, the performance of the policy without power control degrades greatly, as it suffers much from the interference between the two base stations. On the other hand, by using our power control algorithm, the two base stations start to operate in disjoint resource blocks as the distance between them decreases. Hence, the performance of the policy using power control does not suffer too much from interference.

We now consider the Client Association Problem. We consider a system with two macro base stations that are separated by 500m. There are four clients uniformly distributed between them. We consider the performance of our proposed mechanism under various price of energy. Simulation results are shown in Figure \ref{fig:association_performance}. We can see that when the price is small, the performance of the system degrades quickly with price. However, at a price of 0.06, the performance increases, and then degrades with price, but with a smaller slope. This is because, at a price of 0.06, our mechanism determines that it is better to shut down one of the base stations in order to save power and increase energy efficiency. Figure \ref{fig:association_consumption} also shows that the total power consumption of this system decreases by about half at a price of 0.06.

Finally, we present our simulation results for a large scale system. The topology of this system is illustrated in Figure \ref{fig:grid}. We consider a 3000m by 3000m area with 9 macro base stations forming a 3 by 3 grid. In addition, there are 16 micro base stations uniformly distributed in the area $[1000,3000]\times[1000,3000]$. We assume that there are 81 clients uniformly distributed in the area $[0,3000]\times[0,3000]$. Clients within the area of $[0,1000]\times[0,1000]$ have weights $w_i=2$, while all other clients have weights $w_i=1$.

Fig. \ref{fig:full} shows the performance comparison between our proposed and other mechanisms. Fig. \ref{fig:spectrum_efficiency} compares the weighted total throughput, $\sum_i w_ir_i$, and Fig. \ref{fig:energy_efficiency} compares the energy efficiency, defined as $(\sum_i w_ir_i)/$(total power consumption), for the various mechanisms. Our proposed protocol achieves better performance than all other mechanisms, especially when the price of energy is high. Further, as the price of energy increases, our proposed protocol turns some base stations into sleep mode, which results in smaller weighted total throughput but improves energy efficiency. Thus, our proposed protocol can achieve tradeoff between energy efficiency and spectrum efficiency by choosing suitable price of energy.

\section{Conclusion}    \label{section:conclusion}

In this paper, we propose a distributed protocol for self-organizing LTE systems that considers both spectrum efficiency and energy efficiency. This protocol jointly optimizes several important components, including resource block scheduling, power allocation, client association, and the decisions of being in active or sleep mode. The protocol requires small computational and communicational overheads. Further, simulation results show that our proposed protocol achieves much better performance than the existing policy.

\section*{Acknowledgment}

We would like to thank Fran\c{c}ois Baccelli (INRIA-ENS) and Alberto Conte (Alcatel-Lucent) for their valuable discussion and support. Part of this work has been presented at IEEE ICC'12.



\begin{thebibliography}{10}
\providecommand{\url}[1]{#1}
\csname url@samestyle\endcsname
\providecommand{\newblock}{\relax}
\providecommand{\bibinfo}[2]{#2}
\providecommand{\BIBentrySTDinterwordspacing}{\spaceskip=0pt\relax}
\providecommand{\BIBentryALTinterwordstretchfactor}{4}
\providecommand{\BIBentryALTinterwordspacing}{\spaceskip=\fontdimen2\font plus
\BIBentryALTinterwordstretchfactor\fontdimen3\font minus
  \fontdimen4\font\relax}
\providecommand{\BIBforeignlanguage}[2]{{%
\expandafter\ifx\csname l@#1\endcsname\relax
\typeout{** WARNING: IEEEtran.bst: No hyphenation pattern has been}%
\typeout{** loaded for the language `#1'. Using the pattern for}%
\typeout{** the default language instead.}%
\else
\language=\csname l@#1\endcsname
\fi
#2}}
\providecommand{\BIBdecl}{\relax}
\BIBdecl

\bibitem{3GPP-LTE-A}
3rd Generation Partnership~Project. {3GPP LTE-Advanced}.
  http://www.3gpp.org/lte-advanced.

\bibitem{BLTJ10}
\BIBentryALTinterwordspacing
J.~M. Graybeal and K.~Sridhar, ``The evolution of {SON} to extended {SON},''
  \emph{Bell Labs Technical Journal}, vol.~15, no.~3, pp. 5--18, 2010.
  [Online]. Available: \url{http://dx.doi.org/10.1002/bltj.20454}
\BIBentrySTDinterwordspacing

\bibitem{NGMN}
{Next Generation Mobile Networks Group (NGMN)}. http://www.ngmn.org.

\bibitem{GF08}
G.~Fettweis and E.~Zimmermann, ``{ICT} energy consumption - trends and
  challenges,'' in \emph{WPMC}, 2008, pp. 2006--2009.

\bibitem{LTEbook2011}
S.~Sesia, I.~Toufik, and M.~Baker, \emph{LTE - The UMTS Long Term Evolution:
  From Theory to Practice}, 2nd~ed.\hskip 1em plus 0.5em minus 0.4em\relax John
  Wiley \& Son, 2011.

\bibitem{SC_Forum}
{Small Cell Forum}. http://www.smallcellforum.org.

\bibitem{CSChen10}
C.~S. Chen and F.~Baccelli, ``Self-optimization in mobile cellular networks: power control and user association,'' in \emph{Proc. IEEE ICC}, May 2010.

\bibitem{HH10}
H.~Hu, J.~Zhang, X.~Zheng, Y.~Yang, and P.~Wu, ``Self-configuration and
  self-optimization for lte networks,'' \emph{IEEE Communications Magazine},
  no.~2, pp. 94--100, 2010.

\bibitem{Sem11}
S.~Borst, M.~Markakis, and I.~Saniee, ``Distributed power allocation and user
  assignment in {OFDMA} cellular networks,'' in \emph{Proc. Allerton Conference on
  Communication, Control, and Computing}, Sep. 2011.

\bibitem{DLP11}
D.~Lopez-Perez, A.~Ladanyi, A.~Juttner, H.~Rivano, and J.~Zhang, ``Optimization
  method for the joint allocation of modulation schemes, coding rates, resource
  blocks and power in self-organizing lte networks,'' in \emph{Proc.  IEEE INFOCOM},
  2011, pp. 111--115.

\bibitem{Hou11}
I.-H. Hou and P.~Gupta, ``Distributed resource allocation for proportional
  fairness in multi-band wireless systems,'' in \emph{Proc. IEEE ISIT}, Jul. 2011.

\bibitem{Hou12}
I.-H. Hou and C.~S. Chen, ``Self-organized resource allocation in {LTE} systems
  with weighted proportional fairness,'' in \emph{Proc. IEEE ICC}, Jun. 2012, pp. 5348--5353. 

\bibitem{Auer11}
G.~Auer, V.~Giannini, C.~Desset, I.~Godor, P.~Skillermark, M.~Olsson, M.~Imran,
  D.~Sabella, M.~Gonzalez, O.~Blume, and A.~Fehske, ``How much energy is needed
  to run a wireless network?'' \emph{IEEE Wireless Communications}, vol.~18,
  no.~5, pp. 40--49, Oct. 2011.

\bibitem{greenBS11}
S.~Mclaughlin, P.~Grant, J.~Thompson, H.~Haas, D.~Laurenson, C.~Khirallah,
  Y.~Hou, and R.~Wang, ``Techniques for improving cellular radio base station
  energy efficiency,'' \emph{IEEE Wireless Communications}, vol.~18, no.~5, pp.
  10--17, Oct. 2011.

\bibitem{Conte11}
A.~Conte, A.~Feki, L.~Chiaraviglio, D.~Ciullo, M.~Meo, and M.~Marsan, ``Cell
  wilting and blossoming for energy efficiency,'' \emph{IEEE Wireless
  Communications}, vol.~18, no.~5, pp. 50--57, Oct. 2011.

\bibitem{kson11}
K.~Son, H.~Kim, Y.~Yi, and B.~Krishnamachari, ``Base station operation and user
  association mechanisms for energy-delay tradeoffs in green cellular
  networks,'' \emph{IEEE JSAC}, no.~8, pp. 1525--1536, 2011.

\bibitem{SZ09}
S.~Zhou, J.~Gong, Z.~Yang, Z.~Niu, and P.~Yang, ``Green mobile access network
  with dynamic base station energy saving,'' in \emph{Proc. of ACM Mobicom},
  2009.

\bibitem{JG12}
J.~Gong, S.~Zhou, and Z.~Niu, ``A dynamic programming approach for base station
  sleeping in cellular networks,'' \emph{IEICE Transactions on Communications},
  vol. E95-B, no.~2, pp. 551--562, feb. 2012.

\bibitem{YC11}
Y.~Chen, S.~Zhang, S.~Xu, and G.~Li, ``Fundamental trade-offs on green wireless
  networks,'' \emph{IEEE Communications Magazine}, no.~6, pp. 30--37, 2011.

\bibitem{GM09}
G.~Miao, N.~Himayat, Y.~G. Li, and A.~Swami, ``Cross-layer optimization for
  energy-efficient wireless communications: a survey,'' \emph{Wirel. Commun.
  Mob. Comput.}, vol.~9, no.~4, pp. 529--542, Apr. 2009.

\bibitem{survey11}
G.~Li, Z.~Xu, C.~Xiong, C.~Yang, S.~Zhang, Y.~Chen, and S.~Xu,
  ``Energy-efficient wireless communications: tutorial, survey, and open
  issues,'' \emph{IEEE Wireless Communications}, vol.~18, no.~6, pp. 28--35, Dec. 2011.

\bibitem{HK04}
H.~Kim, K.~Kim, Y.~Han, and S.~Yun, ``A proportional fair scheduling for
  multicarrier transmission systems,'' in \emph{Proc. IEEE VTC-Fall}, Sep. 2004, pp. 409--413.

\bibitem{Bazaraa}
M.~S. Bazaraa, H.~D. Sherali, and C.~M. Shetty, \emph{Nonlinear Programming --
  Theory and Algorithms}, 3rd~ed.\hskip 1em plus 0.5em minus 0.4em\relax John
  Wiley \& Son, 2006.

\bibitem{3GPP2010R10UE}
{3GPP~TS~36.101}, ``Evolved universal terrestrial radio access: User equipment radio transmission and reception,'' Tech. Spec. v10.3.0, Jun. 2011.

\end{thebibliography}

\appendices

\section{Proof of Theorem 2}

Theorem \ref{theorem:scheduling:optimal} has shown that the online scheduling policy in Section \ref{section:scheduling} achieves the optimum solution to the Scheduling Problem. We claim that, by setting $\phi_{i,m,z}$ as that derived in Algorithm \ref{algorithm:client:AE} and $\phi_{j,m,z}=\frac{w_j}{\sum_{k\neq i, m(k)=m}w_k}(1-\phi_{i,m,z})$, for all $j\neq i$, $m(j)=m$, the resulting $r_i$ and $\{r_j|j\neq i, m(j)=m\}$ also achieve the optimum solution to the Scheduling Problem.

In the proof of Theorem \ref{theorem:scheduling:optimal}, it has been shown that $\{\phi_{j,m,z}|m(j)=m\}$ maximizes $\sum_{j:m(j)=m}w_j\log r_j$ if and only if $\phi_{j,m,z}\geq 0$, for all $j,z$, $\sum_j \phi_{j,m,z}=1$, for all $m,z$, and $\frac{\partial L}{\partial
\phi_{j,m,z}}=\frac{w_jH_{j,m,z}}{r_j}=\max_{k:m(k)=m}\frac{w_kH_{k,m,z}}{r_k}$, for all $j,z$ such that $\phi_{j,m,z}>0$. By our settings of $\phi_{j,m,z}$, the first two conditions hold, and we only need to verify the last condition.

Sort all resource blocks such that $\frac{\overline{H}_{m,1}}{H_{i,m,1}}\leq \frac{\overline{H}_{m,2}}{H_{i,m,2}}\leq \frac{\overline{H}_{m,3}}{H_{i,m,3}}\leq\dots,$ we consider two possible cases: there exists some $z_0$ such that $0<\phi_{i,m,z_0}<1$, and such $z_0$ does not exist, i.e., $\phi_{i,m,z}\in\{0,1\}$ for all $z$. In the first case, we have
that $\phi_{i,m,z}=1$, for all $z<z_0$, and $\phi_{i,m,z}=0$, for all $z>z_0$. By setting $\phi_{j,m,z}=\frac{w_j}{\sum_{k\neq i, m(k)=m}w_k}(1-\phi_{i,m,z})$, for all $j\neq i$, $m(j)=m$, we have $r_j=\frac{w_j}{\sum_{k:m(k)=m, k\neq i}w_k}\overline{r}_m=\frac{w_j}{w^m_{-i}}\overline{r}_m$. Let $r^*_i$ and $\overline{r}^*_m$ be the values of $r_i$ and $\overline{r}_m$ in the $z_0$-th iteration of the 
\textbf{for} loop in Algorithm \ref{algorithm:client:AE}. As $0<\phi_{i,m,z_0}<1$, lines 13--17 are executed in this iteration, and we have $r_i=r^*_i+\phi_{i,m,z_0}H_{i,m,z_0}$ and $\overline{r}_m=\overline{r}^*_m-\phi_{i,m,z_0}\overline{H}_{m,z_0}$. Moreover, in line 13, the value of $\phi_{i,m,z_0}$ is chosen so that $\frac{w_iH_{i,m,z_0}}{r^*_i+\phi_{i,m,z_0}H_{i,m,z_0}}=\frac{w^m_{-i}\overline{H}_{m,z_0}}{\overline{r}^*_m-\phi_{i,m,z_0}\overline{H}_{m,z_0}}$.
Therefore, $\frac{w_iH_{i,m,z_0}}{r_i}=\frac{w^m_{-i}\overline{H}_{m,z_0}}{\overline{r}_m}=\frac{w_{j}\overline{H}_{m,z_0}}{r_j}$,
for all $j$ such that $m(j)=m$ and $j\neq i$. Thus, the last condition holds for resource block $z_0$. For any resource block $z<z_0$, $\frac{\overline{H}_{m,z}}{H_{i,m,z}}\leq\frac{\overline{H}_{m,z_0}}{H_{i,m,z_0}}$. We then have
$(\frac{w_iH_{i,m,z}}{r_i})/(\frac{w_{j}\overline{H}_{m,z}}{r_j})\geq(\frac{w_iH_{i,m,z_0}}{r_i})/(\frac{w_{j}\overline{H}_{m,z_0}}{r_j})=1$,
and hence $\frac{w_iH_{i,m,z}}{r_i}\geq\frac{w_{j}\overline{H}_{m,z}}{r_j}$, for all $j$ such that $m(j)=m$ and $j\neq i$. As we set $\phi_{j,m,z}=0$ for all $j\neq i$, the last condition holds for all $z<z_0$. Similarly, for any resource block $z>z_0$, $\frac{w_iH_{i,m,z}}{r_i}\leq\frac{w_{j}\overline{H}_{m,z}}{r_j}$, for all $j$ such that $m(j)=m$ and $j\neq i$. As we set $\phi_{i,m,z}=0$, the last condition also holds for all $z>z_0$. In sum, the last condition holds for the case that there exists some $z_0$ such that $0<\phi_{i,m,z_0}<1$.

Next consider the case that $\phi_{i,m,z}\in\{0,1\}$ for all $z$. Let $z_1$ be the smallest integer so that $\phi_{i,m,z_1}=0$. In the $z_1$-th iteration of the FOR loop in Algorithm \ref{algorithm:client:AE}, steps 10--11 are executed, and we have
$\frac{w_iH_{i,m,z}}{r_i}<\frac{w^m_{-i}\overline{H}_{m,z}}{\overline{r}_m}=\frac{w_{j}\overline{H}_{m,z}}{r_j}$,
for all $j$ such that $m(j)=m$ and $j\neq i$. The last condition
then holds for resource block $z_1$. A similar argument as in the previous
paragraph shows that the last condition holds for all $z$.

\vfill

\end{document}